\documentclass[aps]{revtex4}
\usepackage{graphicx}
\usepackage{bm}
\usepackage{graphicx}
\usepackage{amsmath}
\usepackage{mathrsfs}
\usepackage{ulem}
\usepackage{color,xcolor}
\usepackage{amsthm,amsmath,amssymb}
\usepackage[colorlinks, linkcolor=red,anchorcolor=green,citecolor=blue]{hyperref}
\usepackage{cancel}
\begin{document}
\title{Pair structure of heavy tetraquark systems }
\author{Guojun Huang}
\author{Jiaxing Zhao}
\author{Pengfei Zhuang}
\address{Physics Department, Tsinghua University, Beijing 100084, China}

\begin{abstract}
We study the pair description of heavy tetraquark systems $|QQ\bar Q \bar Q\rangle$ in the frame of a non-relativistic potential model. By taking the two heavy quark pairs $(Q\bar Q)$ as colored clusters, the four-quark Schr\"odinger equation is reduced to a two-pair equation, when the inner motion inside the pairs can be neglected. Taking into account all the Casimir scaling potentials between two quarks and using the lattice QCD simulated mixing angle between the two color-singlet states for the tetraquark system, we extracted a detailed pair potential between the two heavy quark pairs.  
\end{abstract}
\maketitle

Among the studies of exotic hadrons which can not be explained as normal mesons and baryons, there are many theoretical works focusing on heavy tetraquark systems $QQ\bar Q\bar Q$ ($Q$ means charm quark $c$ or bottom quark $b$)~\cite{jaffe,ader,zouzou,badalian,brink,weinstein,nielsen,brambilla,berezhnoy,esposito,olsen,hosaka,kang,karliner,guo,debastiani,wang,liu,chen,yang,lu,zhao1}. Recently, a narrow structure around $6.9$~GeV, named as $X(6900)$, is observed by the LHCb Collaboration at colliding energy $\sqrt{s}=7, 8$ and $13$~TeV~\cite{aaij}. This is the first candidate of fully-heavy tetraquarks observed in experiment. 

The molecular picture is an often used mechanism to understand the properties of multi-quark states, for instance the phenomenon that some of the tetraquark states locate below the threshold of the corresponding two mesons~\cite{berezhnoy,karliner,wang1,chen1,anwar1,esposito1,debastiani,bai1,wang2}. The key quantity to control the molecular structure of heavy tetraquark states is the interaction potential between the two molecules. However, if one directly takes a tetraquark state as a meson-meson configuration, there will be no interaction between the two mesons at one-gluon-exchange level, see Ref.\cite{bicudo} and the discussion below. A direct way to introduce meson-meson interaction at quark level is to include multi-gluon exchange. In this paper, we consider a different molecular picture, by taking into account pair-pair interaction at one-gluon-exchange level. We study a colored pair description for general heavy tetraquark systems in the frame of a non-relativistic potential model. By taking the two heavy quark pairs $(Q\bar Q)$ as two colored clusters and neglecting the quark motion inside the pairs, we translate the four quark problem into a two-pair problem and derive the interaction potential between the two pairs.  

The two independent color-singlet states $|s_1\rangle=|(QQ)_{\bar 3_c}(\bar Q\bar Q)_{3_c}\rangle$ and $|s_2\rangle=|(QQ)_{6_c}(\bar Q\bar Q)_{\bar 6_c}\rangle$ form a complete and orthonormal basis in the color space, any tetraquark state can be expanded in terms of them,
\begin{equation}
|QQ\bar Q\bar Q\rangle = \sin \Theta |s_1\rangle +\cos \Theta|s_2\rangle,
\end{equation}
where the mixing angle $\Theta$~\cite{bicudo} between the two color-singlet states characterizes the color dynamics of the tetraquark state. Considering the color structure of the two independent meson-meson states $|m_1\rangle=|(Q_1\bar Q_3)_{1_c}(Q_2\bar Q_4)_{1_c}\rangle$ and $|m_2\rangle=|(Q_1\bar Q_4)_{1_c}(Q_2\bar Q_3)_{1_c}\rangle$, 
\begin{eqnarray}
|m_1\rangle &=& \sqrt{1/3}|s_1\rangle+\sqrt{2/3}|s_2\rangle,\nonumber\\
|m_2\rangle &=& -\sqrt{1/3}|s_1\rangle+\sqrt{2/3}|s_2\rangle,
\end{eqnarray}
a general tetraquark state can be expanded as a linear combination of the two meson-meson states~\cite{bicudo}.
\begin{equation}
|QQ\bar Q\bar Q\rangle = \sqrt{3/4}\left[\left(\cos\Theta/\sqrt 2+\sin\Theta\right)|m_1\rangle +\left(\cos\Theta/\sqrt 2-\sin\Theta\right)|m_2\rangle\right],
\end{equation}
Note that, different from the two color-singlet states $|s_1\rangle$ and $|s_2\rangle$, the two independent meson-meson states are normalized but not orthogonal to each other. It is easy to check that, a general tetraquark state $|QQ\bar Q\bar Q\rangle$ is reduced to the meson-meson state $|m_1\rangle$ at the mixing angle $\Theta=\Theta_0$ with $\tan\Theta_0=1/\sqrt 2$ and the other meson-meson state $|m_2\rangle$ at $\Theta=-\Theta_0$.

An often used method to investigate multi heavy-quark systems is the potential model~\cite{brink,debastiani,wang,chen,yang,lu,zhao1,zhao2}. For a heavy tetraquark system, the wave function $\Psi({\bf r}_1,{\bf r}_2,{\bf r}_3,{\bf r}_4)=\langle {\bf r}_1,{\bf r}_2,{\bf r}_3,{\bf r}_4|QQ\bar Q\bar Q\rangle$ and energy $E$ satisfy the Schr\"odinger equation in coordinate space,
\begin{equation}
\left(\sum_{i=1}^4 {-{\bm \nabla}_i^2\over 2M}+V\right)\Psi=E\Psi
\end{equation}
with quark mass $M$. As a first approximation, we consider only the one-gluon-exchange potential between two quarks, and take the total potential $V({\bf r}_1,{\bf r}_2,{\bf r}_3,{\bf r}_4)$ as a sum of such Casimir scaling potentials~\cite{debbio},
\begin{equation}
\label{vt1}
V =\sum_{i<j}^4{\langle QQ\bar Q\bar Q|\lambda_i^a\otimes\lambda_j^a|QQ\bar Q\bar Q\rangle\over -16/3}V_c(|{\bf r}_{ij}|),
\end{equation}
where the matrix $\lambda^a_i$ is defined as $2T^a$ for quark $i$ and $-2(T^a)^*$ for anti-quark $i$ with $T^a$ being the Gell-Mann matrices, $|{\bf r}_{ij}|=|{\bf r}_j-{\bf r}_i|$ is the distance between the two quarks $i$ and $j$, and $V_c(r)$ is the static Cornell potential
\begin{equation}
V_c(r) =-{\alpha\over r}+\sigma r.
\end{equation}
The two parameters $\alpha$ and $\sigma$ can be fixed by fitting the charmonium masses in vacuum~\cite{zhao2}.
\begin{figure}
	\centering
	\includegraphics[height=5.0cm,width=13cm]{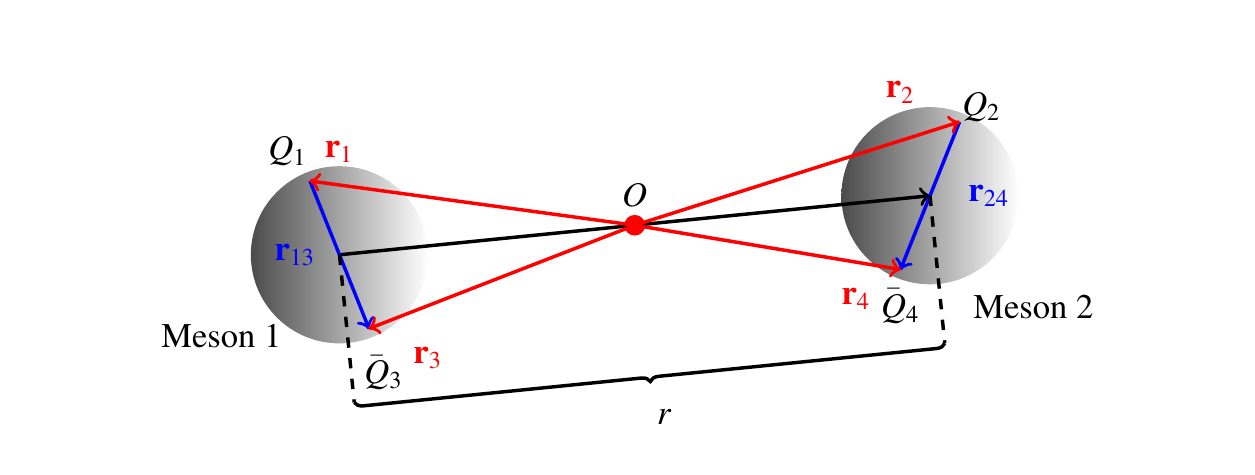}
	\caption{ The relative coordinates of the quarks in a tetraquark system. $O$ is the center-of-mass of the system. }
	\label{fig1}
\end{figure}

Since the Cornell potential depends only on the distance between two interacting quarks, the four-body Schr\"odinger equation can be divided into a center-of-mass part and a relative part $\Psi=\Theta\psi$. The center-of-mass motion is a plane wave $\Theta({\bf R})=e^{i{\bf P}\cdot{\bf R}}$ with total coordinate ${\bf R}=({\bf r}_1+{\bf r}_2+{\bf r}_3+{\bf r}_4)/4$ and total momentum ${\bf P}$, and the relative motion is governed by the potential $V$. There are three independent coordinates for the relative motion. In order to achieve the goal of translating the four-quark problem into a problem of two molecules, we take ${\bf r}_{13}={\bf r}_3-{\bf r}_1, {\bf r}_{24}={\bf r}_4-{\bf r}_2$ and the vector ${\bf r}$ between the two quark pairs $(Q_1\bar Q_3)$ and $(Q_2\bar Q_4)$,
\begin{equation}
{\bf r} = \left({\bf r}_2+{\bf r}_4 -{\bf r}_1 -{\bf r}_3\right)/2
\end{equation}
as the three independent coordinates, see Figure \ref{fig1}. The other coordinate vectors can be expressed in terms of them, 
\begin{eqnarray}
{\bf r}_{12} &=& {\bf r}+({\bf r}_{13}-{\bf r}_{24})/2,\nonumber\\
{\bf r}_{14} &=& {\bf r}+({\bf r}_{13}+{\bf r}_{24})/2,\nonumber\\
{\bf r}_{32} &=& {\bf r}-({\bf r}_{13}+{\bf r}_{24})/2,\nonumber\\
{\bf r}_{34} &=& {\bf r}-({\bf r}_{13}-{\bf r}_{24})/2.
\end{eqnarray}

Note that, the quark pairs $(Q_1\bar Q_3)$ and $(Q_2\bar Q_4)$ here are not mesons, they are not required to be color-singlets. The reason to choose ${\bf r}_{13}, {\bf r}_{24}$ and ${\bf r}$ as the relative coordinates is for comparing the two-pair structure $|(Q_1\bar Q_3)(Q_2\bar Q_4)\rangle$ with the two-meson structure $|m_1\rangle=|(Q_1\bar Q_3)_{1_c}(Q_2\bar Q_4)_{1_c}\rangle$ of the tetraquark system. We will see that the former approaches to the latter in the limit of $r\to\infty$. We can also choose ${\bf r}_{14}, {\bf r}_{23}$ and ${\bf r}$ between the two quark pairs $(Q_1\bar Q_4)$ and $(Q_2\bar Q_3)$ as the relative coordinates for comparing the two-pair structure $|(Q_1\bar Q_4)(Q_2\bar Q_3)\rangle$ with the two-meson structure $|m_2\rangle=|(Q_1\bar Q_4)_{1_c}(Q_2\bar Q_3)_{1_c}\rangle$. For a tetraquark system with the same mass for the four quarks, the two groups of coordinates make no difference in the final result.

Calculating directly the matrix elements in (\ref{vt1}) leads to the total static potential $V({\bf r},{\bf r}_{13},{\bf r}_{24})$ for the tetraquark system,
\begin{eqnarray}
\label{vt2}
V &=& {1\over 8}\left(1-3\cos(2\Theta)\right)\left[V_c(r_{12})+V_c(r_{34})\right]+{1\over 16}\left(7+3\cos(2\Theta)+6\sqrt 2\sin(2\Theta)\right)\left[V_c(r_{13})+V_c(r_{24})\right]\nonumber\\
&& +{1\over 16}\left(7+3\cos(2\Theta)-6\sqrt 2\sin(2\Theta)\right)\left[V_c(r_{14})+V_c(r_{23})\right].
\end{eqnarray}
It is clear to see that, at the specific mixing angle $\Theta=\Theta_0$ or $\Theta=-\Theta_0$ the tetraquark state is reduced to a meson-meson state, the total potential contains only the inner potentials of the two mesons, $V=V_c(r_{13})+V_c(r_{24})$ at $\Theta=\Theta_0$ and $V=V_c(r_{14})+V_c(r_{23})$ at $\Theta=-\Theta_0$, and the interaction between the two mesons totally disappear. In this case, the four-quark system becomes two free mesons, and it is impossible to form a tetraquark state. 

With the known potential, the relative wave function $\psi({\bf r},{\bf r}_{13},{\bf r}_{24})$ and energy $\epsilon$ which characterize the tetraquark structure are determined by the Schr\"odinger equation
\begin{equation}
\label{s1}
\left(-{1\over 2M}\left({\bm \nabla}_r^2+2\left({\bm \nabla}_{13}^2+{\bm \nabla}_{24}^2\right)\right)+V\right)\psi=\epsilon\psi.
\end{equation}
While we have considered the two quark pairs $(Q_1\bar Q_3)$ and $(Q_2\bar Q_4)$ as two colored clusters, this relative equation cannot be factorized as three equations characterizing separately the structures of the two pair states and the two-pair state, since the potential $V$ is a highly mixed function of ${\bf r}_{13}, {\bf r}_{24}$ and ${\bf r}$ and cannot be written as a sum of three independent parts. To construct a cluster description of the tetraquark state, we have to neglect the quark motion inside the pairs, namely we take the variables ${\bf r}_{13}$ and ${\bf r}_{24}$ as the averaged pair sizes $\overline r_{13}$ and $\overline r_{24}$. By subtracting the inner potentials of the two pairs from the tetraquark potential, 
\begin{equation}
V_{pp}({\bf r},\overline r_{13},\overline r_{24})=V({\bf r},\overline r_{13},\overline r_{24})-V_c(\overline r_{13})-V_c(\overline r_{24}),
\end{equation}
the relative Schr\"odinger equation (\ref{s1}) for the four-quark system becomes the equation for the two-pair system,
\begin{equation}
\label{s2}
\left(-{{\bm \nabla}_r^2\over 2M}+V_{pp}\right)\psi_{pp}=\epsilon_{pp}\psi_{pp},
\end{equation}
where $2M$ is the reduced mass of the pair-pair system, $V_{pp}$ is the potential between the two pairs which contains the direct interactions between the two pairs $V_{12}, V_{14}, V_{23}$ and $V_{34}$ and the mixing induced change in the inner potentials, and $\epsilon_{pp}=\epsilon-V_c(\overline r_{13})-V_c(\overline r_{24})$ and $\psi_{pp}({\bf r},\overline r_{13},\overline r_{24})$ are, respectively, the binding energy and relative wave function of the two-pair system.    

In general case, the Schr\"odinger equation (\ref{s2}) cannot be further separated into an angular part and a radial part, since the pair potential $V_{pp}({\bf r},\overline r_{13},\overline r_{24})$ depends still on the relative direction between the two molecules. For the mixing angle $\Theta$, we can take the lattice QCD simulated $\Theta(r,\overline r_{13}, \overline r_{24})$~\cite{bicudo} as a function of the distance between the two pairs with fixed sizes $\overline r_{13}$ and $\overline r_{24}$. To derive a pair potential $V_{pp}$ depending only on the distance $r$, we integrate the tetraquark potential over all the angles,
\begin{equation}
V(r,\overline r_{13},\overline r_{24}) = {1\over 16\pi^2}\int d\Omega_1 d\Omega_2 V(r,\overline{\bf r}_{13},\overline{\bf r}_{24})
\end{equation} 
with $\Omega_1=(\theta_1,\phi_1), \Omega_2=(\theta_2,\phi_2)$ and the definition of 
\begin{eqnarray}
\overline{\bf r}_{13} &=& \overline r_{13}(\sin\theta_1\cos\phi_1,\sin\theta_1\sin\phi_1,\cos\theta_1),\nonumber\\
\overline{\bf r}_{24} &=& \overline r_{24}(\sin\theta_2\cos\phi_2,\sin\theta_2\sin\phi_2,\cos\theta_2)
\end{eqnarray}
for the azimuth angles of $\overline{\bf r}_{13}$ and $\overline{\bf r}_{24}$. After a straightforward calculation, we obtain
\begin{eqnarray}
V(r,\overline r_{13}, \overline r_{24}) &=& \left[{1\over 4}(1-3\cos(2\Theta))+{1\over 8}\left(7+3\cos(2\Theta)-6\sqrt{2}\sin(2\Theta)\right)\right]{\mathcal V}(r,\overline r_{13},\overline r_{24})\nonumber\\	
&&+{1\over 16}\left(7+3\cos(2\Theta)+6\sqrt{2}\sin(2\Theta)\right)\left(V_c(\overline r_{13})+V_c(\overline r_{24})\right)
\end{eqnarray}
with
\begin{eqnarray}
{\mathcal V}\left(r,\overline r_{13},\overline r_{24}\right) &=& -{\alpha\over r}+\sigma r+{\sigma\over 12}{\overline r_{13}^2+\overline r_{24}^2\over r}\nonumber\\
&& +\left\{\begin{aligned}
&\alpha\dfrac{(r_+-r)^2-(r_--r)^2}{\overline r_{13}\overline r_{24}r}-\sigma\frac{(r_+-r)^4-(r_--r)^4}{6\overline r_{13}\overline r_{24}r},&r<{r_{-}}\\
&\alpha\frac{(r_+-r)^2}{\overline r_{13}\overline r_{24}r}-\sigma\frac{(r_+-r)^4}{6\overline r_{13}\overline r_{24}r},&{r_{-}}\leq r<{r_{+}}\\
&0,&r\geq {r_{+}}\\
\end{aligned}\right.
\end{eqnarray}
and $r_\pm =|r_{13}\pm r_{24}|/2$. 

\begin{figure}
\centering
\includegraphics[height=5.0cm,width=12.5cm]{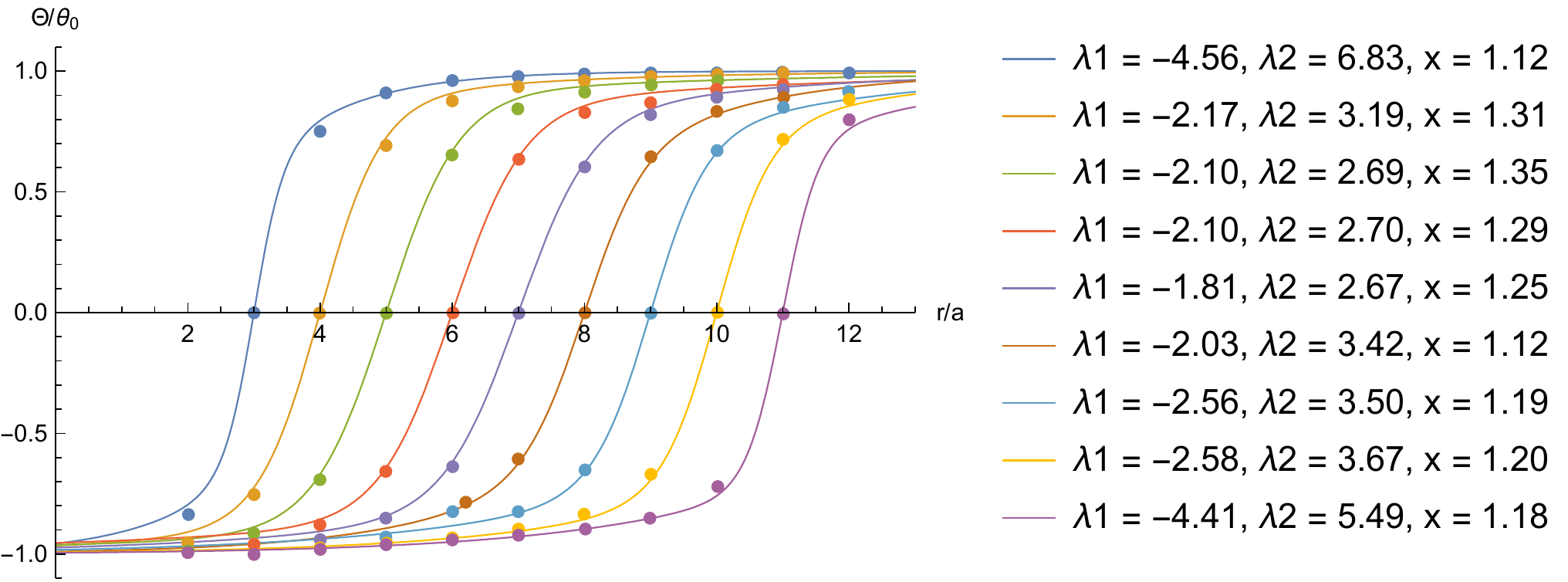}
\caption{The lattice QCD simulated mixing angle $\Theta(r,r_p,r_p)/\Theta_0$ as a function of $r/a$ at $r_p/a=3,4,\cdots,11$ (from left to right)~\cite{bicudo}. The scaled parameters are $\Theta_0=\arctan(1/\sqrt{2})$ and $a=0.069$ fm.  The curves are the fitted results using Eq. (\ref{fitting}) with different values of the parameters $\lambda_1, \lambda_2$ and $x$. }
\label{fig2}
\end{figure}

Now the factor to determine the central pair potential 
\begin{equation}
V_{pp}(r,\overline r_{13},\overline r_{24}) = V(r,\overline r_{13},\overline r_{24})-V_c(\overline r_{13})-V_c(\overline r_{24})
\end{equation}
is the mixing angle $\Theta(r,\overline r_{13},\overline r_{24})$. It is impossible to self-consistently determine it in the frame of potential models. We take here the lattice QCD simulated $\Theta$~\cite{bicudo} as a function of $r$ at fixed pair sizes $\overline r_{13}=\overline r_{24}=r_p$, see Figure \ref{fig2}. In the limit of $r\to\infty$ but with finite pair size $r_p$, $\Theta$ approaches to $\Theta_0$, the tetraquark system becomes two free mesons, and the pair (meson) potential vanishes,  
\begin{equation}
V_{pp}(r\to\infty,r_p,r_p) = 0.
\end{equation}

In the other limit of $\overline r_{13}, \overline r_{24}\to \infty$ but with finite $r$, $\Theta$ approaches to $-\Theta_0$, the tetraquark system becomes again two free mesons, and the averaged pair potential becomes
\begin{equation}
V_{pp}(r,r_p\to\infty,r_p\to\infty) = 2\left(\sigma r-V_c(r_p)\right). 
\end{equation}
The case of $r\to 0$ but with finite $\overline r_{13}$ and $\overline r_{24}$ is close to the second limit, see Figure \ref{fig2}. 

In general case with $0<r<\infty$, the lattice QCD simulated $\Theta$~\cite{bicudo} can be well fitted by   
\begin{equation}
\label{fitting}
{\Theta(r,r_p,r_p)\over \Theta_0} = {\cos^2x\sinh\left[\lambda_1\left(r-r_p\right)\right]+\sin^2x\sinh\left[\lambda_2\left(r-r_p\right)\right]\over\cos^2x\cosh\left[\lambda_1\left(r-r_p\right)\right]+\sin^2x\cosh\left[\lambda_2\left(r-r_p
	\right)\right]}
\end{equation}
with three parameters $\lambda_1, \lambda_2$ and $x$. The lattice data and the fitted lines with different values of the parameters are shown in Figure \ref{fig2}.   

\begin{figure}
	\centering
	\includegraphics[height=5.0cm,width=6.5cm]{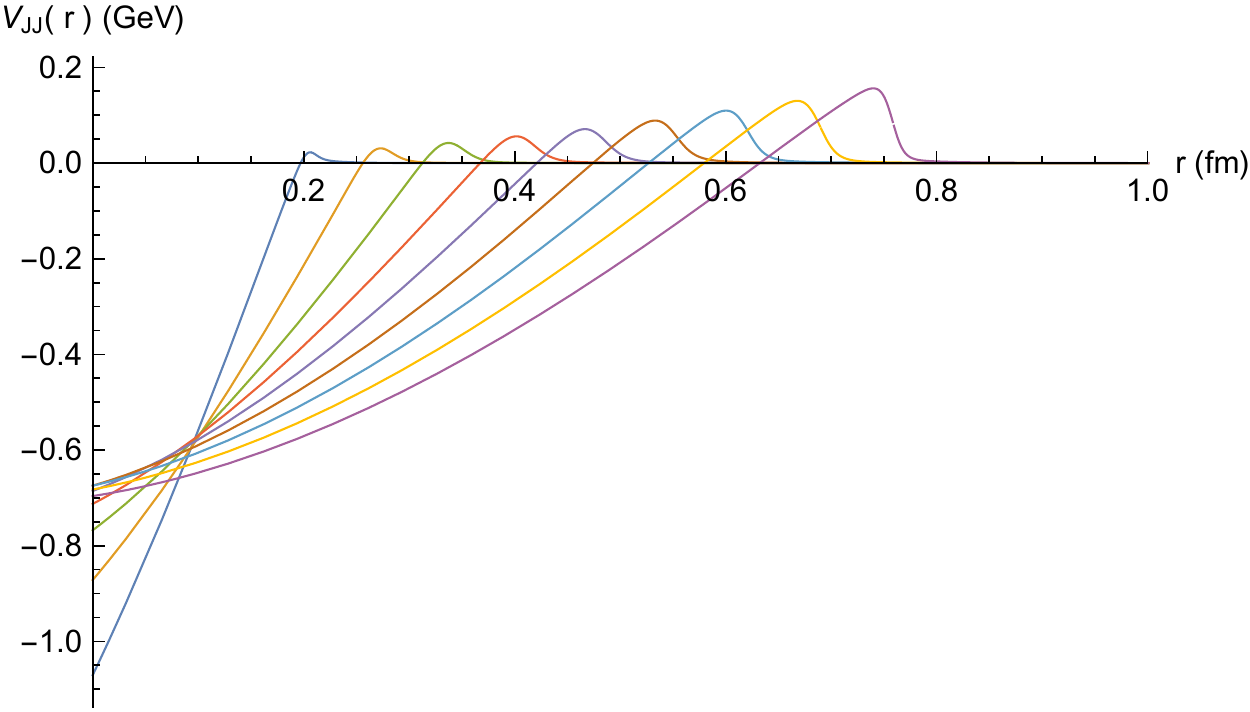}
	\caption{The calculated pair potential as a function of the distance between the two pairs. The 9 lines correspond to different mixing angles taken from Figure \ref{fig2}. }
	\label{fig3}
\end{figure}

Employing the lattice simulated mixing angle $\Theta(r,r_p,r_p)$ with different molecular size $r_p/a=3,4,\cdots,11$, we calculated the pair potential as a function of the distance $r$ between the two pairs. The result is shown in Figure \ref{fig3}, and the corresponding parameters are taken to be $\alpha=0.5$ and $\sigma=0.17$ (GeV)$^2$~\cite{zhao2}. Note that, for tetraquark systems $QQ\bar Q\bar Q$, the pair potential is quark mass independent. Of course, the Schr\"odinger equation or the wave function of the system depends on the mass value. If we take the pair size $r_p$ as two times the $J/\psi$ radius $\sim 0.8$ fm, the pair potential should approach to the $J/\psi-J/\psi$ potential in the limit of large $r$. The potential is attractive at small $r$ which bounds the two pairs together, then becomes continuously repulsive around the pair size $r_p$, and finally approaches to zero when the distance is large enough. The numerical result here is similar to a recent calculation of $J/\psi-J/\psi$ potential where the colorless Pomeron exchange produces the attractive force~\cite{gong}. 

With the central pair potential $V_{pp}(r)$, the relative wave function $\psi_{pp}({\bf r})$ for the two-pair system can be factorized as a spherical harmonic function and a radial function, the latter is controlled by the potential. In above calculations we have taken, for simplicity, the same mass for the four quarks, but the method to effectively reduce a four-quark system to a two-pair system is valid for general tetraquark systems with different quark masses. 
     
In summary, we obtained a pair description of heavy tetraquark systems $|QQ\bar Q\bar Q\rangle$ in the frame of a potential model. Taking the two quark pairs $(Q\bar Q)$ as colored clusters and neglecting the quark motion inside the pairs, the four-body Schr\"odinger equation for the tetraquark system becomes a two-body equation describing the relative motion of the two-pair state. Considering all the Casimir scaling potentials between two heavy quarks and using the lattice QCD simulated mixing angle between the two independent color singlets, we obtained a detailed interaction potential between the two pairs.  

{\bf Acknowledgement}: We thank Professor Lianyi He for helpful discussions in the beginning of this work. The work is supported by Guangdong Major Project of Basic and Applied Basic Research No. 2020B0301030008 and the NSFC under grant Nos. 11890712 and 12075129.


\begin{thebibliography}{99}
\bibitem{jaffe}
R.L.Jaffe, Phys. Rev. {\bf D15}, 267 (1977); Phys. Rev. {\bf D15}, 281(1977).
	
\bibitem{ader}
J.P.Ader, J.M.Richard and P.Taxil, Phys. Rev. {\bf D25}, 2370(1982).

\bibitem{zouzou}
S.Zouzou, B.Silvestre-Brac, C.Gignoux and J.M.Richard, Z. Phys. {\bf C30}, 457(1986).

\bibitem{badalian}
A.M.Badalian, B.L.Ioffe and A.V.Smilga, Nucl. Phys. {\bf B281}, 85(1987).

\bibitem{brink}
D.M.Brink and F.Stancu, Phys. Rev. {\bf D57}, 6778(1998).

\bibitem{weinstein}
J.D.Weinstein, N.Isgur, Phys. Rev. {\bf D41}, 2236(1990).

\bibitem{nielsen}
M.Nielsen, F.S.Navarra, S.H.Lee, Phys. Rept. {\bf 497}, 41(2010).

\bibitem{brambilla}
N.Brambilla $et. al.$ [Heavy Quarkonium Working Group], Eur. Phys. J. {\bf C71}, 1534(2011).

\bibitem{berezhnoy}
A.V.Berezhnoy, A.V.Luchinsky and A.A.Novoselov, Phys. Rev. {\bf D86}, 034004(2012).

\bibitem{esposito}
A.Esposito, A.L.Guerrieri, F.Piccinini, A.Pilloni, A.D.Polosa, Int. J. Mod. Phys. {\bf A30}, 1530002(2014).

\bibitem{olsen}
S.L.Olsen, Front. Phys. {\bf 10}, 101401(2015).

\bibitem{hosaka}
A.Hosaka, T.Iijima, K.Miyabayashi, Y.Sakai and S.Yasui, PTEP{\bf 2016}, 062C01(2016).

\bibitem{kang}
X.W.Kang and J.A.Oller, Eur. Phys. J. {\bf C77}, 399(2017).

\bibitem{karliner}
M.Karliner, S.Nussinov and J.L.Rosner, Phys. Rev. {\bf D95}, 034011(2017).

\bibitem{guo}
F.K.Guo, C.Hanhart, U.G.Meißner, Q.Wang, Q.Zhao and B.S.Zou, Rev. Mod. Phys. {\bf 90}, 015004(2018).

\bibitem{debastiani}
V.R.Debastiani and F.S.Navarra, Chin. Phys. {\bf C43}, 013105(2019).

\bibitem{wang}
G.J.Wang, L.Meng and S.L.Zhu, Phys. Rev. {\bf D100}, 096013(2019).

\bibitem{liu}
M.S.Liu, Q.F.Lü, X.H.Zhong and Q.Zhao, Phys. Rev. {\bf D100}, 016006(2019).

\bibitem{chen}
X.Chen, [arXiv:2001.06755 [hep-ph]].

\bibitem{yang}
G.Yang, J.Ping, L.He and Q.Wang, [arXiv:2006.13756 [hep-ph]].

\bibitem{lu}
Q.F.Lü, D.Y.Chen and Y.B.Dong, [arXiv:2006.14445 [hep-ph]].

\bibitem{zhao1}
J.Zhao, S.Shi and P.Zhuang, Phys. Rev. {\bf D}, (2020). 
		
\bibitem{aaij}
R.Aaij \textit{et al.} [LHCb Collaboration], [arXiv:2006.16957 [hep-ex]].
	
\bibitem{wang1}
Z.Wang, Eur. Phys. J. {\bf C77}, 432(2017).	

\bibitem{chen1}
W.Chen, H.Chen, X.Liu, T.Steele and S.Zhu, Phys. Lett. {\bf B773}, 247(2017).

\bibitem{anwar1}
M.Anwar, J.Ferretti, F.Guo, E.Santopinto and B.Zou, Eur. Phys. J. {\bf C78}, 647(2018).	

\bibitem{esposito1}
A.Esposito and A.Polosa, Eur Phys J. {\bf C78}, 782(2018).	
	
\bibitem{bai1}
Y.Bai, S.Lu and J.Osborne, Phys. Lett. {\bf B798}, 134930(2019).

\bibitem{wang2}
Z.Wang and Z.Di, Acta Phys. Polon. {\bf B50}, 1335(2019).
	
\bibitem{bicudo}
P.Bicudo, M.Cardoso, O.Oliveira and P.J.Silva, Phys. Rev. {\bf D96}, 074508(2017). 

\bibitem{zhao2}
J.Zhao, K.Zhou, S.Chen and P.Zhuang, Prog. Part. Nucl. Phys. {\bf 114}, 103801(2020).

\bibitem{debbio}
L.Del Debbio, M.Faber, J.Greensite, S.Olejnik,  Phys. Rev. ${\bf D53}$, 5891(1996).

\bibitem{gong}
C.Gong, M.C.Du, B.Zhou, Q.Zhao and X.H.Zhong, arXiv:2011.11374.
	
\end{thebibliography}
\end{document}